\documentclass[%
reprint,
superscriptaddress,
 amsmath,amssymb,
 aps,prl,
]{revtex4-1}

\usepackage[utf8]{inputenc} 
\usepackage{todonotes}
\usepackage{graphicx}%
\usepackage{dcolumn}%
\usepackage{bm}%
\usepackage[normalem]{ulem}
\usepackage{microtype}
\usepackage{xcolor}
\usepackage{amsmath,amssymb,amsfonts,mathtools,cancel}
\usepackage{multirow}
\usepackage{comment}

\usepackage[paperwidth=210mm,
            paperheight=297mm,
            left=20mm,
            top=10mm,
            textwidth=170mm,
            marginparsep=3mm,
            marginparwidth=30mm,
            textheight=730pt,
            footskip=50pt]
           {geometry}

\newcommand{\kb}{k_{\rm B}}

\newcommand{\avg}[1]{\ensuremath{\langle{#1}\rangle}}

\newcommand{\br}{\ensuremath{\textbf{r}}}
\newcommand{\bk}{\ensuremath{\textbf{k}}}

\newcommand{\figLabel}[1]{\textbf{\textsf{\MakeLowercase{#1}}}}  %
\newcommand{\figLabelCapt}[1]{\textbf{\textsf{\MakeLowercase{#1}}}}  %

\newcommand{\refSub}[2]{\hyperref[#2]{\ref{#2}\figLabelCapt{#1}}}

\newcommand{\secref}[2]{\hyperref[#2]{#1}}

\newcommand{\subfigimg}[3][,]{%
  \setbox1=\hbox{\includegraphics[#1]{#3}}%
  \leavevmode\rlap{\usebox1}%
  \rlap{\hspace*{-2pt}\raisebox{\dimexpr\ht1-2\baselineskip}{#2}}%
  \phantom{\usebox1}%
}

\begin{document}

\preprint{APS/123-QED}

\title{
Computing chemical potentials of adsorbed or confined fluids
}%

\author{Rochus Schmid}
\affiliation{Computational Materials Chemistry Group, Faculty of Chemistry and Biochemistry, Ruhr-Universität Bochum, Universitätsstr. 150, 44801 Bochum, Germany}

\author{Bingqing Cheng}%
\email{bingqing.cheng@ist.ac.at}
\affiliation{The Institute of Science and Technology Austria, Am Campus 1, 3400 Klosterneuburg, Austria}%

\date{\today}%

\begin{abstract}
The chemical potential of adsorbed or confined fluids provides insight into their unique thermodynamic properties and determines
adsorption isotherms.
However, 
it is often difficult to compute this quantity from atomistic simulations using existing statistical mechanical methods.
We introduce a computational framework that utilizes static structure factors, thermodynamic integration and free energy perturbation, for calculating the absolute chemical potential of fluids.
For demonstration, we apply the method to compute the adsorption isotherms of carbon dioxide in a metal-organic framework (MOF) and water in carbon nanotubes.
\end{abstract}

\maketitle

When a fluid is confined to a small volume, such as inside a nanotube or an ion channel, or is adsorbed in a porous material, its chemical potential can be significantly different from that of the bulk fluid~\cite{gelb1999phase}.
The chemical potential directly determines the
adsorption isotherm,
which is the relationship between the amount of a gas or liquid that is adsorbed and the pressure or concentration of the gas or liquid in the surrounding reservoir. 
This relationship is important for porous materials, because it determines their capacity to store and separate fluids~\cite{gelb1999phase}, and can also have a significant impact on their physical and chemical properties~\cite{fumagalli2018anomalously}.
The adsorption isotherm also helps understand the phase behavior~\cite{kapil2022first,algara2015square} and the transport phenomena of confined fluids~\cite{kavokine2022fluctuation}.

Computing chemical potentials from atomistic simulations can be challenging,
especially for confined or adsorbed fluids.
Many methods
~\cite{sanz2007solubility,paluch2010method,lisal2005molecular,moucka2011molecular,perego2016chemical,joung2008determination,Li2017,Li2018,Vinutha2021}
have caveats and only work for a subset of systems:
Monte Carlo particle insertion and removal ~\cite{allen2012computer,smit1989calculation} can have the problems of low insertion probability or numerical convergence issues~\cite{perego2016chemical}.
Thermodynamic integration (TI) or overlapping distribution method~\cite{sanz2007solubility,paluch2010method}
may have singularity problems at the end points of the integration~\cite{Li2017,Li2018}.
Moreover, pressure, which enters the TI expression along an isotherm,
is ill-defined for confined systems~\cite{varnik2000molecular,Shi2023} or heterogeneous fluids~\cite{marchio2018pressure}. 

The recent S0 method~\cite{Cheng2022computing} brings a new perspective for chemical potentials of mixtures. 
This method utilizes the thermodynamic relationship between 
the particle number fluctuations and the derivatives of the chemical potentials with respect to the molar concentration,
and only uses the static structure factors computed from equilibrium molecular dynamics (MD) simulations at different mixture fractions.
However, there are a couple of leaps to be made to apply the method for the chemical potentials of adsorbed or confined fluids.
First,
the S0 method evaluates the relative chemical potential as a function of concentration $c$ for each phase, and
one needs to establish the relation between these relative values of different phases.
Second, the method was developed for fluid mixtures,
so one needs to extend it to other situations.

Here we introduce a statistical mechanical method (S0-TIFEP) to compute the chemical potentials and the adsorption
isotherms of adsorbed or confined fluids.
We then benchmark the applicability of the method for the simple system of an  
argon fluid in a MOF,
a carbon dioxide fluid in MOF-5, an archetypical rigid metal organic framework with a cubic pore structure~\cite{yaghi1999MOF5},
and water in single-walled carbon nanotubes (CNTs).

\section{Theory}

\begin{figure}
\includegraphics[width=0.45\textwidth]{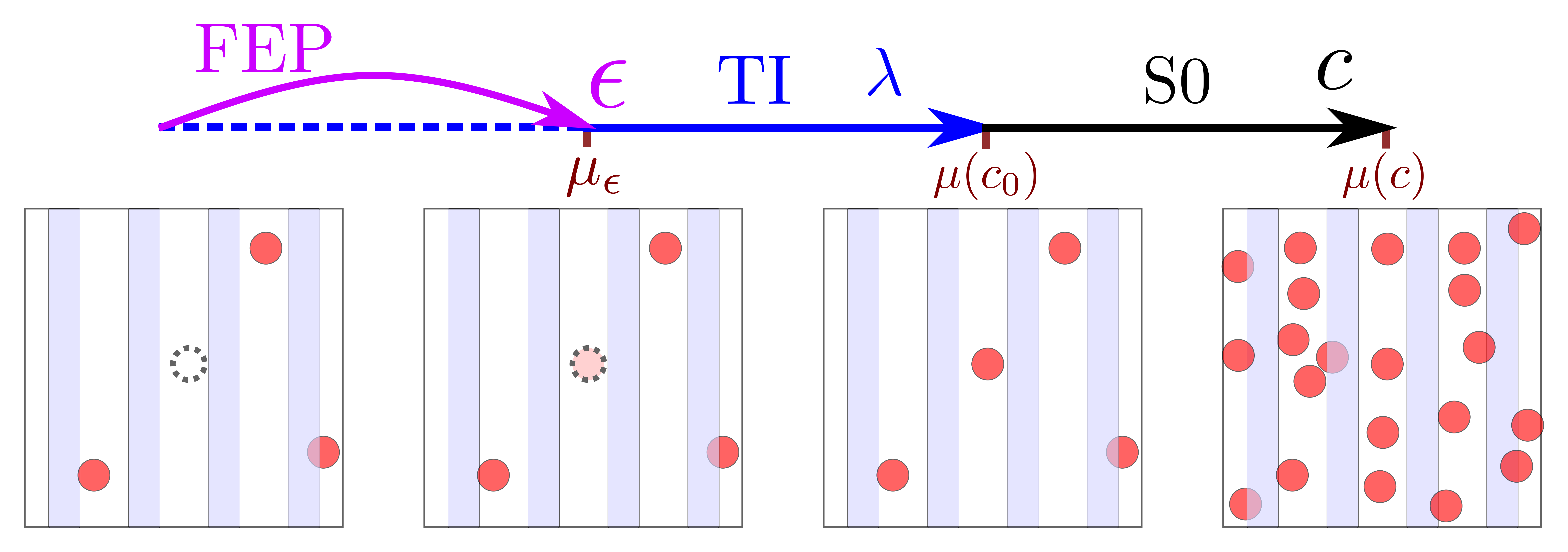}
\caption{
The thermodynamic framework for computing the absolute chemical potential of adsorbed or confined fluid.
The TIFEP method for calculating the excess chemical potential at concentration $c_0$ is shown by the
purple and blue arrows:
first use FEP to add $\epsilon$ fraction of a molecule, 
and then reversibly switching this fraction of a molecule to a fully-interacting one in TI.
Black arrow indicates using S0 method to compute the chemical potential at different concentrations $c$.
}
\label{fig:method}
\end{figure}

The general workflow is illustrated in Fig.~\ref{fig:method}.
For the adsorbed phase, 
one first computes the absolute chemical potential at a certain low loading, using a combination of thermodynamic integration and free energy perturbation (TIFEP),
and then determines the chemical potentials at other concentrations using the S0 method.
For the pure phase, the same approach applies,
although sometimes one can compute $\mu(c)$ directly starting from the dilute limit using the S0 method or TI.
In what follows we describe each step of the workflow.

\paragraph{\textbf{Absolute chemical potential}}
The absolute chemical potential, $\mu$, is defined as the free energy difference between a particle that is fully-interacting with its surroundings and an isolated particle.
The $\mu$ at a certain particle concentration, $c$, can be split into an ideal and an excess part:
\begin{equation}
    \mu = \mu_{id}(c) + \mu_{ex},
\end{equation}
where the ideal part can be analytically expressed as $\mu_{id}(c) =\mu_0 + \kb T \ln(c/c_0)$ for large systems.
To determine $\mu_{ex}$,
one can use the conventional particle insertion method for certain cases,
but the TIFEP approach (the purple and the blue arrows in Fig.~\ref{fig:method}) is more robust and statistically efficient.
The TI is performed along a reversible path between the physical system and a reference system over a switching parameter $\lambda$~\cite{frenkel1984new,cheng2018computing}.
The parameterized Hamiltonian is 
\begin{equation}
    \mathcal{H}(\lambda)=(1-\lambda)\mathcal{H}_{no-inter}+\lambda \mathcal{H}
\end{equation}
where $\mathcal{H}$ is the actual Hamiltonian, and $\mathcal{H}_{no-inter}$ is for the reference system with no interaction between one ghost molecule and the rest of the system. 
In $\mathcal{H}_{no-inter}$, the intramolecular energy and forces of the ghost molecule, and all the interactions between the rest of the system stay the same as $\mathcal{H}$. 
The excess free energy of the extra molecule can then be evaluated using
\begin{equation}
  \mu^{ex}=\int_0^1 d \lambda 
  \avg{U-U_{no-inter}}_{\lambda},
  \label{eq:ti-l}
\end{equation}
where $\avg{\ldots}_{\lambda}$ denotes the ensemble average over NVT or NPT simulations using the Hamiltonian $H(\lambda)$.
However, the integrand in Eqn.~\eqref{eq:ti-l} is divergent at $\lambda=0$,
because when the ghost molecule and the rest of the system are not interacting at all, the atoms can overlap and cause extremely large energy differences $U-U_{no-inter}$.
This problematic bit is indicated by the blue dashed arrow in Fig.~\ref{fig:method}.
To circumvent this issue, one can perform free energy perturbation (FEP) at the end point (purple arrow in Fig.~\ref{fig:method}), i.e.
\begin{equation}
    \Delta \mu_\epsilon^{ex} =
    - k_B T\ln\avg{
    \exp\left[-\dfrac{\epsilon(U-U_{no-inter})}{k_B T}\right]}_{\lambda=0},
    \label{eq:gcut}
\end{equation}
where $\epsilon$ is a small number, which we typically set to about 0.05 though it can be flexible. 
In principle, one can also evaluate $\Delta \mu_\epsilon^{ex}$ with $H(\epsilon)$ using a backward FEP,
but this is less statistically efficient
because in the exponential average of the backward FEP low-occurrence outliers will yield extremely large contributions~\cite{cheng2014direct}.
The excess chemical potential is thus
\begin{equation}
  \mu^{ex}=\Delta \mu_\epsilon^{ex} + \int_\epsilon^1 d \lambda 
  \avg{U-U_{no-inter}}_{\lambda}.
  \label{eq:tifep}
\end{equation}

\paragraph{\textbf{Concentration-dependent $\mu$ }}

One can use the S0 method to compute the concentration dependency in $\mu$,
as schematically shown by the black arrow in Fig.~\ref{fig:method}.
To extend the S0 method in Ref.~\cite{Cheng2022computing} to a fluid that is adsorbed or confined in a porous medium that is without frozen long-range disorder,
there are two views:
The first view is to regard the fluid and the medium as two components,
and the latter component is an immobile single molecule with zero associated density fluctuations.
The second view is to only analyze the density fluctuations of the fluid, while treating the medium as a background that only provides an external potential energy field.
The two views result in the same thermodynamic relations,
and here we focus on the second one and introduce the single-component formulation of the S0 method.

For a single-component system, or a single-component system adsorbed in a homogeneous material without arrested defects,
the particle number fluctuations in the grand-canonical ensemble (constant-$\mu$VT) are related to the chemical potential by
\begin{equation}
        \dfrac{\avg{(N-\avg{N})^2}_{\mu VT}}{\avg{N}_{\mu V T}} =
        \dfrac{\kb T}{\avg{N}} \left( \dfrac{\partial \avg{N}}{\partial \mu}\right)_{V,T}.
        \label{eq:fluc-1}
\end{equation}
The structure factor is related to the particle number fluctuations via
\begin{equation}
    S^0 \equiv \lim_{\bk \rightarrow 0} S(\bk) =
    \dfrac{\avg{(N-\avg{N})^2}_{\mu VT}}{\avg{N}_{\mu V T}},
    \label{eq:fluc-s}
\end{equation}
where
\begin{equation}
        S(\bk) = \dfrac{1}{N}\avg{\widetilde{\rho}(\bk,t)\widetilde{\rho}(-\bk,t)},
\end{equation}
with 
\begin{equation}
\widetilde{\rho}(\bk,t) = 
\int_V d\br \rho(\br,t) \exp(i \bk \cdot \br)
= \sum_{i=1}^{N} \exp(i \bk \cdot \br_{i}(t)).
\label{eq:rhokt}
\end{equation}
In practice, one can compute $S(\bk)$ at small $\bk$ from NVT or NPT MD simulations, and extrapolate to the $\bk \rightarrow 0$ case to get $S^0$~\cite{Cheng2022computing}.

Taking the molar concentration as $c=N/V = \avg{N}_{\mu V T}/V$,
and 
combining Eqn.~\eqref{eq:fluc-1} with Eqn.~\eqref{eq:fluc-s}, one gets
\begin{equation}
    \left(\frac {\partial \mu}{\partial c}\right)_{T}
    =\dfrac{k_BT}{c S^0}.
    \label{eq:mu-s}
\end{equation}
Importantly, Eqn.~\eqref{eq:mu-s} works across phase transitions and the coexistence region for large systems:
if the overall $c$ of the system is in between the molar concentrations of two coexisting phases, 
both phases will appear simultaneously with macroscopic phase boundaries,
which will cause $S^0$ to diverge so $\frac {\partial \mu}{\partial c}$ under coexistence will be correctly predicted to vanish.

As the key equation of the method, one can obtain $\mu(c)$ using numerical integration with respect to $c$:
\begin{equation}
    \mu(c) = \mu(c_0) + \kb T \ln(\dfrac{c}{c^0}) 
    + \kb T \int_{\ln c^0 }^{\ln c} d \ln(c)
   \left[ \dfrac{1}{S^0} -1 \right].
    \label{eq:S-integral}
\end{equation}
Notice that a change of the variable $y=\ln(c)$ is done to
reduce the numerical errors in the integration.

Eqn.~\eqref{eq:S-integral} and Eqn.~\eqref{eq:tifep} can then be combined to obtain the chemical potentials of the adsorbed or pure fluids.
Moreover, for a pure fluid,
 $S^0$ also determines the pressure via
\begin{equation}
    \left(\frac {\partial P}{\partial c}\right)_{T}
    = \dfrac{k_BT}{S^0},
    \label{eq:P-S0}
\end{equation}
which becomes the ideal gas law $P=ck_B T$ when $S^0=1$.
Eqn.~\eqref{eq:P-S0} can be useful when the equation of state (EOS) is difficult to compute directly.
Furthermore, from $\dfrac{\partial\mu}{\partial c} \dfrac{\partial c}{\partial P}$ it is easy to verify that
\begin{equation}
    \left(\frac {\partial \mu}{\partial P}\right)_{T}
    = \dfrac{1}{c},
    \label{eq:ti-P}
\end{equation}
which is a useful relationship for computing the $\mu$ of pure liquids using thermodynamic integration along isotherm.
However, unlike the S0 route, Eqn.~\eqref{eq:ti-P} is not applicable to systems with ill-defined pressures, or across phase transitions.

\section{Results}

\begin{figure}
\includegraphics[width=0.49\textwidth]{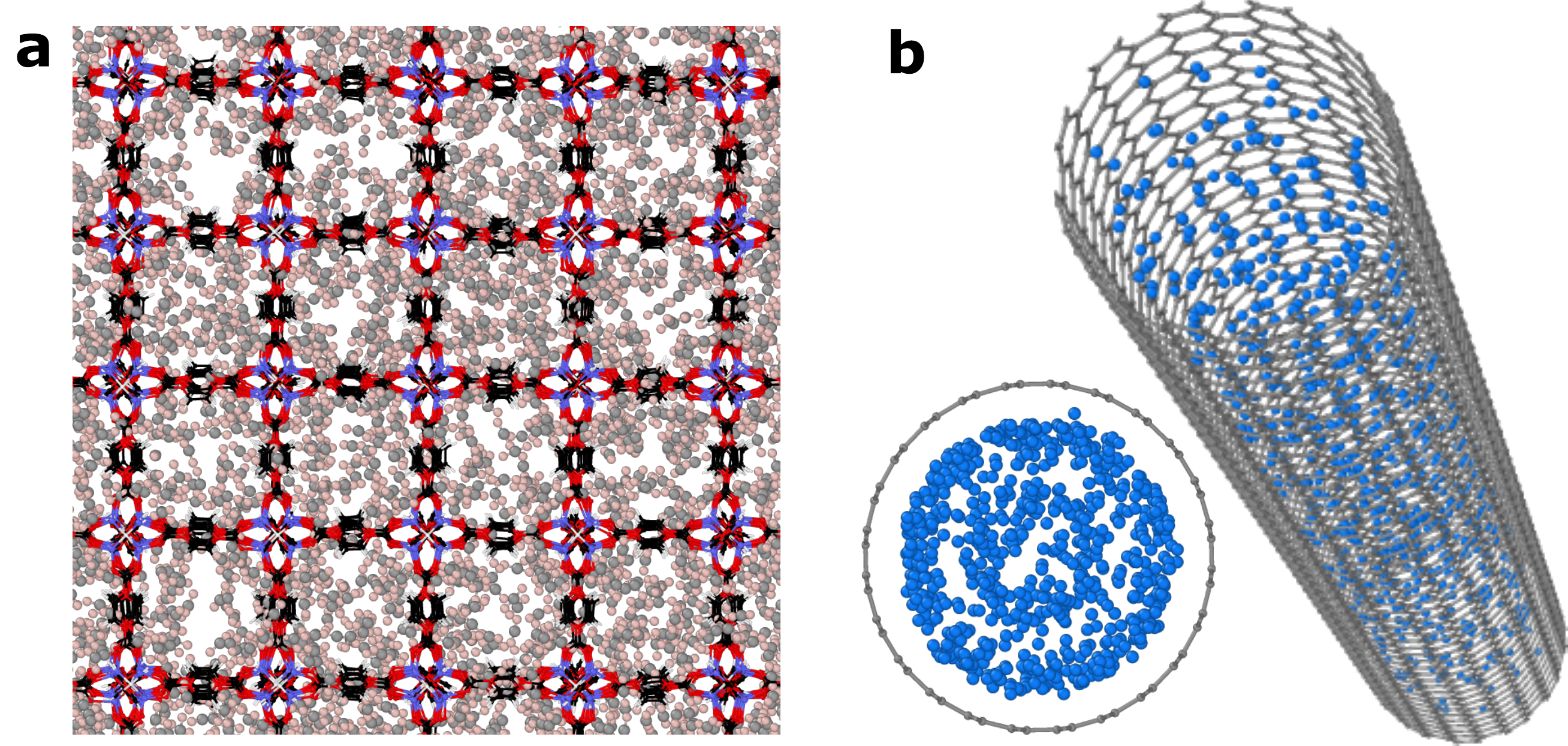}
\caption{
Systems of study:
carbon dioxide in MOF-5 (\figLabelCapt{a}),
and 
water in a single-walled CNT (\figLabelCapt{b}).
}
\label{fig:systems}
\end{figure}

We first validate our method on argon adsorption in MOF-5 at 370~K,
since a number of free energy methods, such as TI, FEP and coexistence simulations, work for this monoatomic gas at high $T$ and allow a benchmark.
Indeed, as detailed in the SI, all methods including S0-TIFEP give consistent results.
We then showcase our method on 
the two systems shown in Fig.~\ref{fig:systems}:
carbon dioxide in MOF-5,
and water in single-walled CNTs.

\paragraph{\textbf{Carbon dioxide in MOF-5}}

The archetypical MOF-5 (or IRMOF-1)~\cite{yaghi1999MOF5} is formed by a Zn$_4$O inorganic building unit, connected by the ditopic terephthalate linker in an octahedral fashion. 
Its interaction with gas has been intensively studied both experimentally and theoretically~\cite{RevMOFAds_Zhou2009,RevMOFads_Snurr2012,RevMOFAds_Yaghi2013,zhou2012introduction,sumida2012carbon}, particularly for the adsorption of carbon dioxide (CO$_2$)~\cite{Walton2007}.
We simulated CO$_2$ in MOF-5 and gas-phase CO$_2$ at 218~K and 300~K using the fully flexible, first-principles-derived forcefield MOF-FF~\cite{PSSB2013MOFFF,Keupp2021FD}. Details on the forcefield parametrization and benchmarks can be found in the SI.
We first computed the absolute $\mu$ for infinitely-dilute CO$_2$ in MOF-5 using the TIFEP method,
by switching from a fully-interacting MOF-5 and one CO$_2$ molecule system to a reference state that has the MOF-5 and a non-interacting ghost CO$_2$ molecule,
using a $1\times1\times1$ MOF-5 cell.
We ran independent simulations on a dense grid of $\lambda$, with $\epsilon=0.01$.
Crucially, a combination of a stochastic velocity rescaling thermostat~\cite{Bussi2007} and a weak local Langevin thermostat was used,
to ensure sufficient equilibration of the ghost molecule.

We then computed the concentration dependence of the 
chemical potentials for CO$_2$ in MOF-5 as well as for the gas phase CO$_2$ using the S0 method.
We performed LAMMPS~\cite{Plimpton1995} NPT simulations at 1~bar for a $6\times6\times6$ supercell of 
 MOF-5 loaded with different amounts of CO$_2$, and NVT simulations of the pure carbon dioxide gas with a similar cell size at a range of $c$.
We used a timestep of 1 fs, since the resulting S0 is consistent with the ones at a smaller time step of 0.1 fs.

\begin{figure}
  \centering
  \begin{tabular}{@{}p{0.45\textwidth}@{\quad}p{0.45\textwidth}@{}}
  \vspace{-0.2cm}
      \subfigimg[width=\linewidth]{\figLabel{a}}{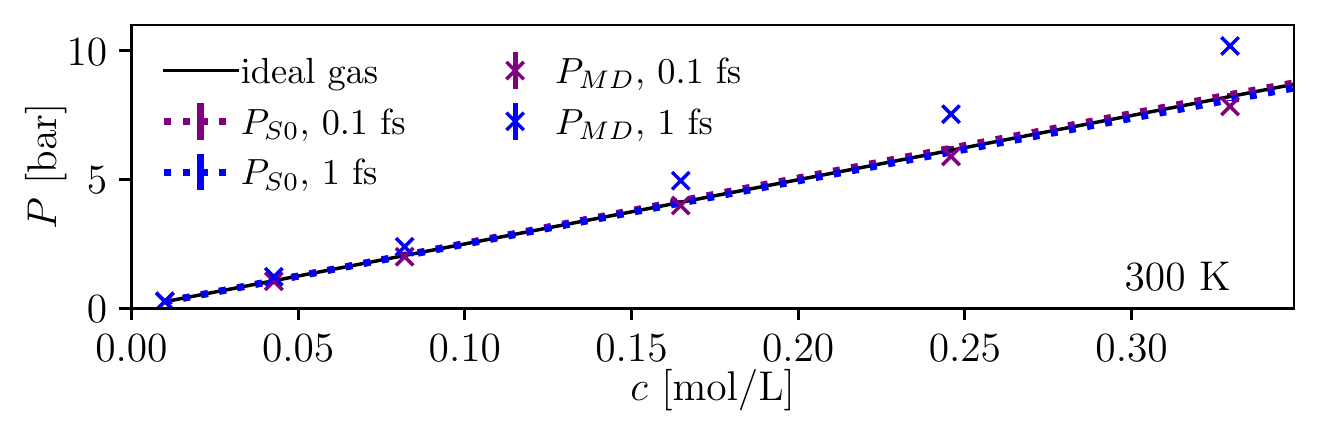} 
      \vspace{-0.2cm}
    \subfigimg[width=\linewidth]{\figLabel{b}}{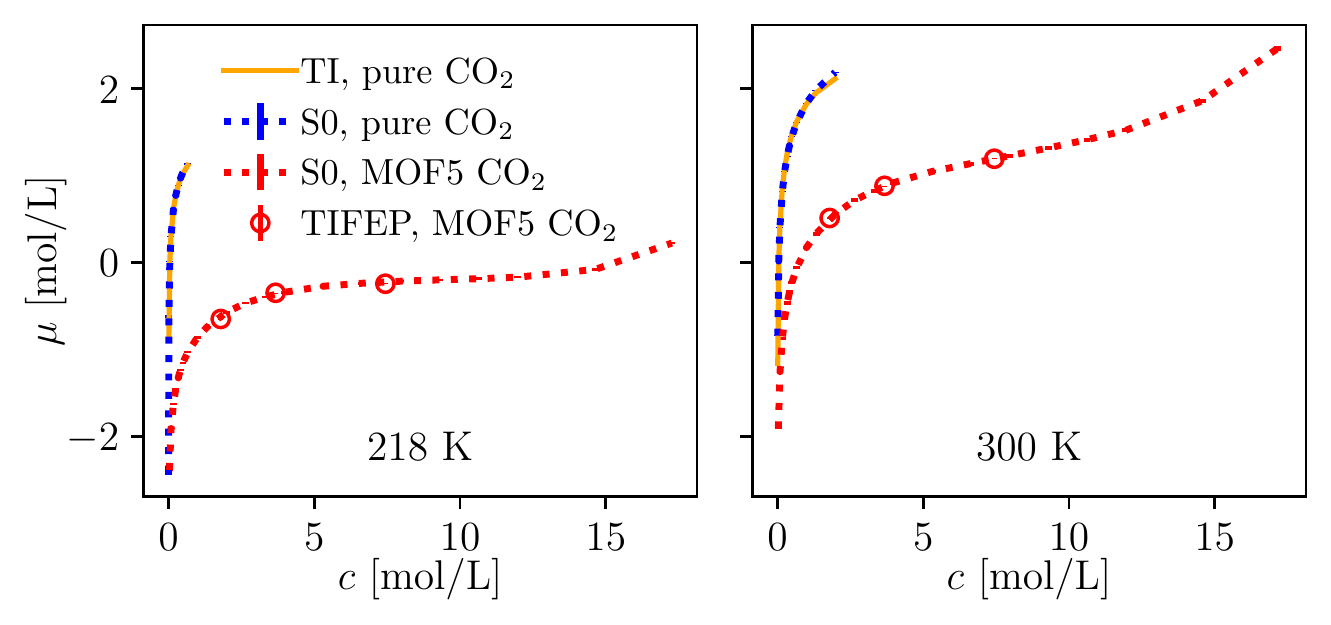} 
    \vspace{-0.2cm}
    \subfigimg[width=\linewidth]{\figLabel{c}}{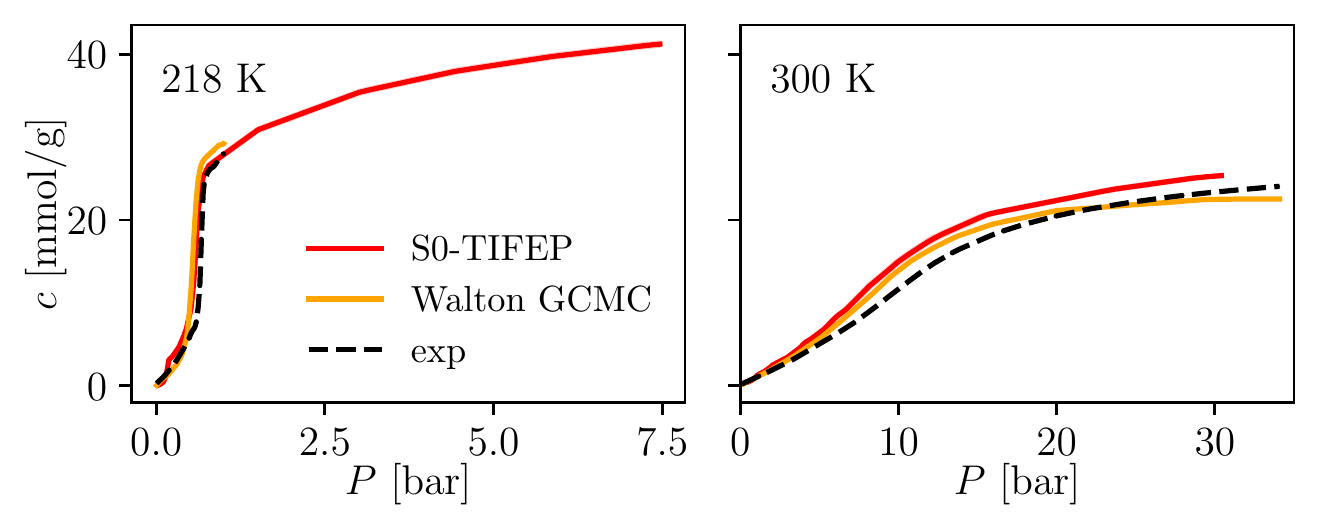}
  \end{tabular}
    \caption{
    \figLabelCapt{a}: The equation of state for gas carbon dioxide.
    \figLabelCapt{b}: The chemical potentials of carbon dioxide molecules in the gas phase and adsorbed in the MOF-5, computed using different TIFEP directly (hollow symbols) and using S0 together with TIFEP at a low concentration (dashed curves). 
    \figLabelCapt{c}: The absolute adsorption isotherm for carbon dioxide in MOF-5. The shaded red areas indicate the uncertainty in our S0-TIFEP calculations, but they are too narrow to be seen. Experimental and GCMC results from Ref. ~\cite{Walton2007} are shown for comparison.}
    \label{fig:mof5-co2}
\end{figure}

To determine the EOS for the gas CO$_2$,
we used 
Eqn.~\eqref{eq:P-S0} employing S0,
which is efficient and insensitive to the timestep,
as shown in Fig.~\ref{fig:mof5-co2}a.
Alternatively, one can take the average $P$ in the NVT simulations,
but for CO$_2$ this has caveats:
the intramolecular bonds are quite stiff and a tiny time step is needed to properly integrate the equation of motion, because the coupling between the van der Waals interactions and the bonds is so weak that it is difficult to equilibrate the latter properly. 
One thus has to use a small timestep of 0.1~fs, and again use a combination of a stochastic velocity rescaling thermostat~\cite{Bussi2007} and a weak local Langevin thermostat.
Overall, the EOS of gas CO$_2$ at low concentrations is rather close to the ideal gas state.

The chemical potentials for the pure and the adsorbed carbon dioxide are plotted in Fig.~\ref{fig:mof5-co2}b.
For the pure gas, the S0 results (blue dashed curve) agree well with the TI along isotherm (Eqn.~\eqref{eq:ti-P}).
For CO$_2$ in MOF-5, we performed additional independent TIFEP calculations at higher concentrations (hollow symbols),
which agree well with the values obtained using S0 together with TIFEP at the dilute concentration (red dashed curves).

To compute the absolute adsorption
isotherm, one relates the concentrations of bulk and adsorbed carbon dioxide that correspond to the same absolute $\mu$ 
and plots the adsorbed concentration against the pressure of the bulk phase.
In Fig.~\ref{fig:mof5-co2}c the equilibrium concentrations of CO$_2$ adsorbed in MOF-5 at 218~K and 300~K are plotted as a function of the pressure of the gas reservoir.
MOF-5 can adsorb a large amount of carbon dioxide.
As the pressure increases, the material becomes saturated and the capability of carbon dioxide adsorption decreases. For both temperatures we observe an inflection and a type V isotherm,
which is more pronounced at the lower temperature. 
The same type V was observed previously experimentally and in grand canonical Monte Carlo (GCMC) simulations~\cite{Walton2007}.
A type V isotherm is usually considered to be quite rare in microporous adsorption.
Our simulated isotherms well-capture
the overall shape of the
experimental curves at both temperatures,
although the simulated curve at 300~K is slightly higher which may be attributed to the defect-free MOF-5 in simulations or the forcefield assumed. 
Differences between our results and the previous GCMC isotherms~\cite{Walton2007} may be due to the different forcefields since they used a rigid MOF-5 with the TraPPE potential for CO$_2$~\cite{TraPPE_CO2}.

\paragraph{\textbf{Water in CNTs}}

Carbon nanotubes have unique physical and chemical properties,
and in particular, their capacity to store or convey water~\cite{alexiadis2008molecular}.
We performed simulations for (8:8) and (12:12) single-walled CNT and water using a simple mW model~\cite{molinero2009water} at 273~K and 500~K,
with the parameterization same as the graphitic and water system in Ref.~\cite{davies2021routes}.
Note that the forcefields for CNT-water are a controversial topic~\cite{alexiadis2008molecular,werder2003water}, and we do not intend to provide a definitive picture for this complex system, but only to demonstrate our methodology.

The TIFEP part was computed from small systems, and the S0 was performed on large systems, as described in the SI.
For water in CNT, $S^0$ was evaluated from $S(\bk)$ with $\bk$ parallel to the direction of the CNT.
Here, the $c$ was defined by the total number of water molecules and the entire volume enclosed by the radius of the CNT.
One can define $c$ differently to consider the volume occupied by the carbon atoms at the wall,
but this will only introduce a constant scaling factor, $\alpha c$, with $\alpha$ close to 1.
Eqn.~\eqref{eq:mu-s} indicates that the scaling $\alpha$ does not affect the values of $\mu$.

\begin{figure}
  \centering
  \begin{tabular}{@{}p{0.45\textwidth}@{\quad}p{0.45\textwidth}@{}}

    \subfigimg[width=\linewidth]{\figLabel{a}}{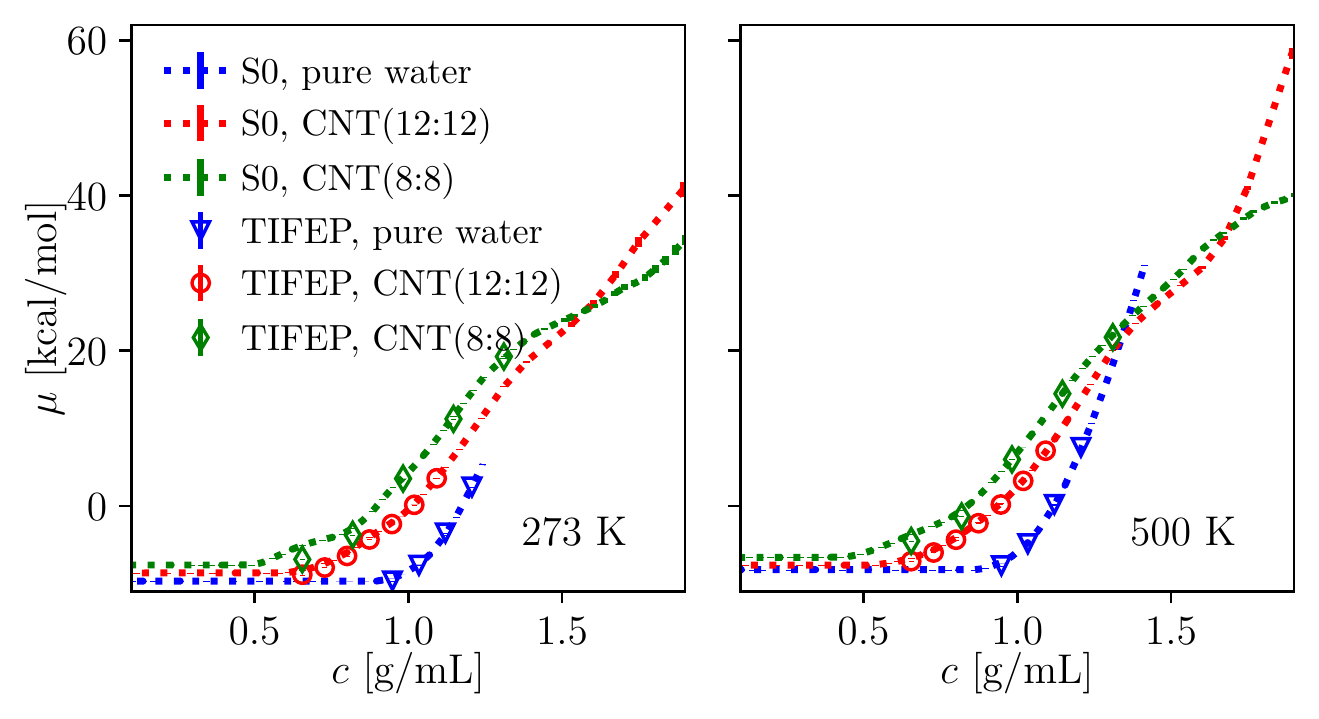}

    \subfigimg[width=\linewidth]{\figLabel{b}}{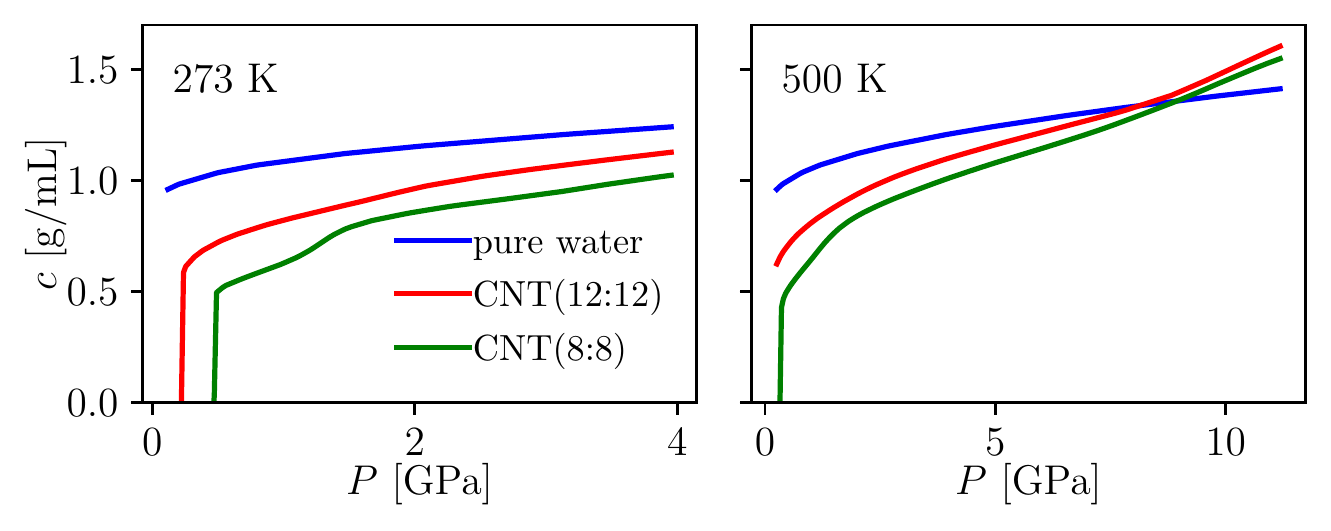}
  \end{tabular}
    \caption{
    \figLabelCapt{a}: Chemical potential of pure water and water in CNTs computed using different methods. 
    \figLabelCapt{b}: The
adsorption isotherm for water.}
    \label{fig:cnt12-water}
\end{figure}

For pure water with $\rho$ less than about 1~g/mL, CNT(12:12)-water with $\rho$ less than about 0.6~g/mL, and CNT(8:8)-water with $\rho$ less than about 0.6~g/mL, the systems show liquid-vapor coexistence.
At these conditions, the presence of the interfaces has an influence on the pressure via the Laplace equation~\cite{marchio2018pressure}, making the actual pressure difficult to compute.
Moreover, pressure is not well-defined for nanoconfined fluids as sizes and stress tensors along the small dimensions are ambiguous~\cite{varnik2000molecular,Shi2023}. 
These make it problematic to obtain $\mu$ using TI along isotherms (Eqn.~\eqref{eq:ti-P}).
In contrast, the S0 method has no problem handling coexistence: $S^0$ approaches infinity, giving zero $\frac {\partial P}{\partial c}$ and zero $\frac {\partial \mu}{\partial c}$ for coexistence systems.
The nanoconfinement is taken into account by only analyzing $S^0$ along the unconfined dimension.

The chemical potentials for bulk and confined water are plotted in Fig.~\ref{fig:cnt12-water}a.
The values obtained using S0 together with TIFEP at a low concentration (dashed curves) agree well 
with independent TIFEP at higher concentrations (hollow symbols).
For systems in the liquid-vapor coexistence region, $\mu$ is constant,
showing the capability of the S0 method to cross phase boundaries.
At low concentrations, bulk water has lower $\mu$ than confined water,
but the confined water has a slower chemical potential increase and becomes more thermodynamically favorable at higher densities.
Bulk water exhibits a solid-liquid phase transition at about 1.2 g/ML at 273~K and 1.5 g/mL at 500~K in the MD simulations,
while such transitions are absent in the CNT-water.

The adsorption
isotherm is shown in Fig.~\ref{fig:cnt12-water}b,
with the pressure of the bulk water plotted on the x-axis.
Note that the absorption isotherm here is based on the thermodynamic equilibrium argument,
which avoids the adsorption-desorption hysteresis loops in the direct simulations~\cite{Striolo2005}. 
CNT~(8:8) has little water adsorption at zero pressure, followed by a steep increase of water intake around 0.5~GPa at both 273~K and 500~K.
Such adsorption behavior exhibits 
the typical features of the type V adsorption isotherms,
and is in agreement with previous simulation results~\cite{Striolo2005}.
CNT~(12:12) follows a similar trend at 273~K, but with a lower threshold pressure of 0.25~GPa.
At high P of more than 10 GPa, the CNT~(8:8) is most efficient in adsorbing water.

\section{Conclusions}
Here we introduce a thermodynamic framework that treats the confined and adsorbed fluids on the same footing.
The framework computes the absolute chemical potentials, and determines the pressure in an efficient and unambiguous way.
We envisage
this S0-TIFEP method will be applied to many technologically important systems for which the chemical potentials of confined or adsorbed fluids can be useful:
Understanding the thermodynamics is key to explore the exotic phases of matter under confinement~\cite{kapil2022first,algara2015square};
The chemical potentials of molecules adsorbed in porous materials are crucial for the performance of fuel cells~\cite{wang2020porous},
gas storage and separation~\cite{gelb1999phase},
reactivity of fluids adsorbed in porous catalysts~\cite{slater2015function},
oil recovery systems~\cite{adebajo2003porous},
and environmental pollution management~\cite{liu2022metal}.

\textbf{Acknowledgments}
BC acknowledges resources provided by the Cambridge Tier-2 system operated by the University of Cambridge Research Computing Service funded by EPSRC Tier-2 capital grant EP/P020259/1. 

\textbf{Data availability statement}
Additional simulation details are provided in the SI.
All PYTHON scripts and simulation input files generated for the study are in the
SI repository (url to be inserted upon acceptance of the paper).


\begin{thebibliography}{44}%
\makeatletter
\providecommand \@ifxundefined [1]{%
 \@ifx{#1\undefined}
}%
\providecommand \@ifnum [1]{%
 \ifnum #1\expandafter \@firstoftwo
 \else \expandafter \@secondoftwo
 \fi
}%
\providecommand \@ifx [1]{%
 \ifx #1\expandafter \@firstoftwo
 \else \expandafter \@secondoftwo
 \fi
}%
\providecommand \natexlab [1]{#1}%
\providecommand \enquote  [1]{``#1''}%
\providecommand \bibnamefont  [1]{#1}%
\providecommand \bibfnamefont [1]{#1}%
\providecommand \citenamefont [1]{#1}%
\providecommand \href@noop [0]{\@secondoftwo}%
\providecommand \href [0]{\begingroup \@sanitize@url \@href}%
\providecommand \@href[1]{\@@startlink{#1}\@@href}%
\providecommand \@@href[1]{\endgroup#1\@@endlink}%
\providecommand \@sanitize@url [0]{\catcode `\\12\catcode `\$12\catcode
  `\&12\catcode `\#12\catcode `\^12\catcode `\_12\catcode `\%12\relax}%
\providecommand \@@startlink[1]{}%
\providecommand \@@endlink[0]{}%
\providecommand \url  [0]{\begingroup\@sanitize@url \@url }%
\providecommand \@url [1]{\endgroup\@href {#1}{\urlprefix }}%
\providecommand \urlprefix  [0]{URL }%
\providecommand \Eprint [0]{\href }%
\providecommand \doibase [0]{http://dx.doi.org/}%
\providecommand \selectlanguage [0]{\@gobble}%
\providecommand \bibinfo  [0]{\@secondoftwo}%
\providecommand \bibfield  [0]{\@secondoftwo}%
\providecommand \translation [1]{[#1]}%
\providecommand \BibitemOpen [0]{}%
\providecommand \bibitemStop [0]{}%
\providecommand \bibitemNoStop [0]{.\EOS\space}%
\providecommand \EOS [0]{\spacefactor3000\relax}%
\providecommand \BibitemShut  [1]{\csname bibitem#1\endcsname}%
\let\auto@bib@innerbib\@empty
\bibitem [{\citenamefont {Gelb}\ \emph {et~al.}(1999)\citenamefont {Gelb},
  \citenamefont {Gubbins}, \citenamefont {Radhakrishnan},\ and\ \citenamefont
  {Sliwinska-Bartkowiak}}]{gelb1999phase}%
  \BibitemOpen
  \bibfield  {author} {\bibinfo {author} {\bibfnamefont {L.~D.}\ \bibnamefont
  {Gelb}}, \bibinfo {author} {\bibfnamefont {K.}~\bibnamefont {Gubbins}},
  \bibinfo {author} {\bibfnamefont {R.}~\bibnamefont {Radhakrishnan}}, \ and\
  \bibinfo {author} {\bibfnamefont {M.}~\bibnamefont {Sliwinska-Bartkowiak}},\
  }\href@noop {} {\bibfield  {journal} {\bibinfo  {journal} {Reports on
  Progress in Physics}\ }\textbf {\bibinfo {volume} {62}},\ \bibinfo {pages}
  {1573} (\bibinfo {year} {1999})}\BibitemShut {NoStop}%
\bibitem [{\citenamefont {Fumagalli}\ \emph {et~al.}(2018)\citenamefont
  {Fumagalli}, \citenamefont {Esfandiar}, \citenamefont {Fabregas},
  \citenamefont {Hu}, \citenamefont {Ares}, \citenamefont {Janardanan},
  \citenamefont {Yang}, \citenamefont {Radha}, \citenamefont {Taniguchi},
  \citenamefont {Watanabe} \emph {et~al.}}]{fumagalli2018anomalously}%
  \BibitemOpen
  \bibfield  {author} {\bibinfo {author} {\bibfnamefont {L.}~\bibnamefont
  {Fumagalli}}, \bibinfo {author} {\bibfnamefont {A.}~\bibnamefont
  {Esfandiar}}, \bibinfo {author} {\bibfnamefont {R.}~\bibnamefont {Fabregas}},
  \bibinfo {author} {\bibfnamefont {S.}~\bibnamefont {Hu}}, \bibinfo {author}
  {\bibfnamefont {P.}~\bibnamefont {Ares}}, \bibinfo {author} {\bibfnamefont
  {A.}~\bibnamefont {Janardanan}}, \bibinfo {author} {\bibfnamefont
  {Q.}~\bibnamefont {Yang}}, \bibinfo {author} {\bibfnamefont {B.}~\bibnamefont
  {Radha}}, \bibinfo {author} {\bibfnamefont {T.}~\bibnamefont {Taniguchi}},
  \bibinfo {author} {\bibfnamefont {K.}~\bibnamefont {Watanabe}},  \emph
  {et~al.},\ }\href@noop {} {\bibfield  {journal} {\bibinfo  {journal}
  {Science}\ }\textbf {\bibinfo {volume} {360}},\ \bibinfo {pages} {1339}
  (\bibinfo {year} {2018})}\BibitemShut {NoStop}%
\bibitem [{\citenamefont {Kapil}\ \emph {et~al.}(2022)\citenamefont {Kapil},
  \citenamefont {Schran}, \citenamefont {Zen}, \citenamefont {Chen},
  \citenamefont {Pickard},\ and\ \citenamefont {Michaelides}}]{kapil2022first}%
  \BibitemOpen
  \bibfield  {author} {\bibinfo {author} {\bibfnamefont {V.}~\bibnamefont
  {Kapil}}, \bibinfo {author} {\bibfnamefont {C.}~\bibnamefont {Schran}},
  \bibinfo {author} {\bibfnamefont {A.}~\bibnamefont {Zen}}, \bibinfo {author}
  {\bibfnamefont {J.}~\bibnamefont {Chen}}, \bibinfo {author} {\bibfnamefont
  {C.~J.}\ \bibnamefont {Pickard}}, \ and\ \bibinfo {author} {\bibfnamefont
  {A.}~\bibnamefont {Michaelides}},\ }\href@noop {} {\bibfield  {journal}
  {\bibinfo  {journal} {Nature}\ }\textbf {\bibinfo {volume} {609}},\ \bibinfo
  {pages} {512} (\bibinfo {year} {2022})}\BibitemShut {NoStop}%
\bibitem [{\citenamefont {Algara-Siller}\ \emph {et~al.}(2015)\citenamefont
  {Algara-Siller}, \citenamefont {Lehtinen}, \citenamefont {Wang},
  \citenamefont {Nair}, \citenamefont {Kaiser}, \citenamefont {Wu},
  \citenamefont {Geim},\ and\ \citenamefont {Grigorieva}}]{algara2015square}%
  \BibitemOpen
  \bibfield  {author} {\bibinfo {author} {\bibfnamefont {G.}~\bibnamefont
  {Algara-Siller}}, \bibinfo {author} {\bibfnamefont {O.}~\bibnamefont
  {Lehtinen}}, \bibinfo {author} {\bibfnamefont {F.}~\bibnamefont {Wang}},
  \bibinfo {author} {\bibfnamefont {R.~R.}\ \bibnamefont {Nair}}, \bibinfo
  {author} {\bibfnamefont {U.}~\bibnamefont {Kaiser}}, \bibinfo {author}
  {\bibfnamefont {H.}~\bibnamefont {Wu}}, \bibinfo {author} {\bibfnamefont
  {A.~K.}\ \bibnamefont {Geim}}, \ and\ \bibinfo {author} {\bibfnamefont
  {I.~V.}\ \bibnamefont {Grigorieva}},\ }\href@noop {} {\bibfield  {journal}
  {\bibinfo  {journal} {Nature}\ }\textbf {\bibinfo {volume} {519}},\ \bibinfo
  {pages} {443} (\bibinfo {year} {2015})}\BibitemShut {NoStop}%
\bibitem [{\citenamefont {Kavokine}\ \emph {et~al.}(2022)\citenamefont
  {Kavokine}, \citenamefont {Bocquet},\ and\ \citenamefont
  {Bocquet}}]{kavokine2022fluctuation}%
  \BibitemOpen
  \bibfield  {author} {\bibinfo {author} {\bibfnamefont {N.}~\bibnamefont
  {Kavokine}}, \bibinfo {author} {\bibfnamefont {M.-L.}\ \bibnamefont
  {Bocquet}}, \ and\ \bibinfo {author} {\bibfnamefont {L.}~\bibnamefont
  {Bocquet}},\ }\href@noop {} {\bibfield  {journal} {\bibinfo  {journal}
  {Nature}\ }\textbf {\bibinfo {volume} {602}},\ \bibinfo {pages} {84}
  (\bibinfo {year} {2022})}\BibitemShut {NoStop}%
\bibitem [{\citenamefont {Sanz}\ and\ \citenamefont
  {Vega}(2007)}]{sanz2007solubility}%
  \BibitemOpen
  \bibfield  {author} {\bibinfo {author} {\bibfnamefont {E.}~\bibnamefont
  {Sanz}}\ and\ \bibinfo {author} {\bibfnamefont {C.}~\bibnamefont {Vega}},\
  }\href@noop {} {\bibfield  {journal} {\bibinfo  {journal} {The Journal of
  chemical physics}\ }\textbf {\bibinfo {volume} {126}},\ \bibinfo {pages}
  {014507} (\bibinfo {year} {2007})}\BibitemShut {NoStop}%
\bibitem [{\citenamefont {Paluch}\ \emph {et~al.}(2010)\citenamefont {Paluch},
  \citenamefont {Jayaraman}, \citenamefont {Shah},\ and\ \citenamefont
  {Maginn}}]{paluch2010method}%
  \BibitemOpen
  \bibfield  {author} {\bibinfo {author} {\bibfnamefont {A.~S.}\ \bibnamefont
  {Paluch}}, \bibinfo {author} {\bibfnamefont {S.}~\bibnamefont {Jayaraman}},
  \bibinfo {author} {\bibfnamefont {J.~K.}\ \bibnamefont {Shah}}, \ and\
  \bibinfo {author} {\bibfnamefont {E.~J.}\ \bibnamefont {Maginn}},\
  }\href@noop {} {\bibfield  {journal} {\bibinfo  {journal} {The Journal of
  chemical physics}\ }\textbf {\bibinfo {volume} {133}},\ \bibinfo {pages}
  {124504} (\bibinfo {year} {2010})}\BibitemShut {NoStop}%
\bibitem [{\citenamefont {L{\'\i}sal}\ \emph {et~al.}(2005)\citenamefont
  {L{\'\i}sal}, \citenamefont {Smith},\ and\ \citenamefont
  {Kolafa}}]{lisal2005molecular}%
  \BibitemOpen
  \bibfield  {author} {\bibinfo {author} {\bibfnamefont {M.}~\bibnamefont
  {L{\'\i}sal}}, \bibinfo {author} {\bibfnamefont {W.~R.}\ \bibnamefont
  {Smith}}, \ and\ \bibinfo {author} {\bibfnamefont {J.}~\bibnamefont
  {Kolafa}},\ }\href@noop {} {\bibfield  {journal} {\bibinfo  {journal} {The
  Journal of Physical Chemistry B}\ }\textbf {\bibinfo {volume} {109}},\
  \bibinfo {pages} {12956} (\bibinfo {year} {2005})}\BibitemShut {NoStop}%
\bibitem [{\citenamefont {Moucka}\ \emph {et~al.}(2011)\citenamefont {Moucka},
  \citenamefont {L{\'\i}sal}, \citenamefont {skvor}, \citenamefont
  {Jirsak}, \citenamefont {Nezbeda},\ and\ \citenamefont
  {Smith}}]{moucka2011molecular}%
  \BibitemOpen
  \bibfield  {author} {\bibinfo {author} {\bibfnamefont {F.}~\bibnamefont
  {Moucka}}, \bibinfo {author} {\bibfnamefont {M.}~\bibnamefont {L{\'\i}sal}},
  \bibinfo {author} {\bibfnamefont {J.}~\bibnamefont {skvor}}, \bibinfo
  {author} {\bibfnamefont {J.}~\bibnamefont {Jirs{\'a}k}}, \bibinfo {author}
  {\bibfnamefont {I.}~\bibnamefont {Nezbeda}}, \ and\ \bibinfo {author}
  {\bibfnamefont {W. R.}\ \bibnamefont {Smith}},\ }\href@noop {} {\bibfield
  {journal} {\bibinfo  {journal} {The Journal of Physical Chemistry B}\
  }\textbf {\bibinfo {volume} {115}},\ \bibinfo {pages} {7849} (\bibinfo {year}
  {2011})}\BibitemShut {NoStop}%
\bibitem [{\citenamefont {Perego}\ \emph {et~al.}(2016)\citenamefont {Perego},
  \citenamefont {Giberti},\ and\ \citenamefont
  {Parrinello}}]{perego2016chemical}%
  \BibitemOpen
  \bibfield  {author} {\bibinfo {author} {\bibfnamefont {C.}~\bibnamefont
  {Perego}}, \bibinfo {author} {\bibfnamefont {F.}~\bibnamefont {Giberti}}, \
  and\ \bibinfo {author} {\bibfnamefont {M.}~\bibnamefont {Parrinello}},\
  }\href@noop {} {\bibfield  {journal} {\bibinfo  {journal} {The European
  Physical Journal Special Topics}\ }\textbf {\bibinfo {volume} {225}},\
  \bibinfo {pages} {1621} (\bibinfo {year} {2016})}\BibitemShut {NoStop}%
\bibitem [{\citenamefont {Joung}\ and\ \citenamefont
  {Cheatham~III}(2008)}]{joung2008determination}%
  \BibitemOpen
  \bibfield  {author} {\bibinfo {author} {\bibfnamefont {I.~S.}\ \bibnamefont
  {Joung}}\ and\ \bibinfo {author} {\bibfnamefont {T.~E.}\ \bibnamefont
  {Cheatham~III}},\ }\href@noop {} {\bibfield  {journal} {\bibinfo  {journal}
  {The journal of physical chemistry B}\ }\textbf {\bibinfo {volume} {112}},\
  \bibinfo {pages} {9020} (\bibinfo {year} {2008})}\BibitemShut {NoStop}%
\bibitem [{\citenamefont {Li}\ \emph {et~al.}(2017)\citenamefont {Li},
  \citenamefont {Totton},\ and\ \citenamefont {Frenkel}}]{Li2017}%
  \BibitemOpen
  \bibfield  {author} {\bibinfo {author} {\bibfnamefont {L.}~\bibnamefont
  {Li}}, \bibinfo {author} {\bibfnamefont {T.}~\bibnamefont {Totton}}, \ and\
  \bibinfo {author} {\bibfnamefont {D.}~\bibnamefont {Frenkel}},\ }\href
  {\doibase 10.1063/1.4983754} {\bibfield  {journal} {\bibinfo  {journal} {The
  Journal of Chemical Physics}\ }\textbf {\bibinfo {volume} {146}},\ \bibinfo
  {pages} {214110} (\bibinfo {year} {2017})}\BibitemShut {NoStop}%
\bibitem [{\citenamefont {Li}\ \emph {et~al.}(2018)\citenamefont {Li},
  \citenamefont {Totton},\ and\ \citenamefont {Frenkel}}]{Li2018}%
  \BibitemOpen
  \bibfield  {author} {\bibinfo {author} {\bibfnamefont {L.}~\bibnamefont
  {Li}}, \bibinfo {author} {\bibfnamefont {T.}~\bibnamefont {Totton}}, \ and\
  \bibinfo {author} {\bibfnamefont {D.}~\bibnamefont {Frenkel}},\ }\href
  {\doibase 10.1063/1.5040366} {\bibfield  {journal} {\bibinfo  {journal} {The
  Journal of Chemical Physics}\ }\textbf {\bibinfo {volume} {149}},\ \bibinfo
  {pages} {054102} (\bibinfo {year} {2018})}\BibitemShut {NoStop}%
\bibitem [{\citenamefont {Vinutha}\ and\ \citenamefont
  {Frenkel}(2021)}]{Vinutha2021}%
  \BibitemOpen
  \bibfield  {author} {\bibinfo {author} {\bibfnamefont {H.~A.}\ \bibnamefont
  {Vinutha}}\ and\ \bibinfo {author} {\bibfnamefont {D.}~\bibnamefont
  {Frenkel}},\ }\href {\doibase 10.1063/5.0038955} {\bibfield  {journal}
  {\bibinfo  {journal} {The Journal of Chemical Physics}\ }\textbf {\bibinfo
  {volume} {154}},\ \bibinfo {pages} {124502} (\bibinfo {year}
  {2021})}\BibitemShut {NoStop}%
\bibitem [{\citenamefont {Allen}\ and\ \citenamefont
  {Tildesley}(2012)}]{allen2012computer}%
  \BibitemOpen
  \bibfield  {author} {\bibinfo {author} {\bibfnamefont {M.~P.}\ \bibnamefont
  {Allen}}\ and\ \bibinfo {author} {\bibfnamefont {D.~J.}\ \bibnamefont
  {Tildesley}},\ }\href@noop {} {\emph {\bibinfo {title} {Computer simulation
  in chemical physics}}},\ Vol.\ \bibinfo {volume} {397}\ (\bibinfo
  {publisher} {Springer Science \& Business Media},\ \bibinfo {year} {2012})\
  pp.\ \bibinfo {pages} {108--111}\BibitemShut {NoStop}%
\bibitem [{\citenamefont {Smit}\ and\ \citenamefont
  {Frenkel}(1989)}]{smit1989calculation}%
  \BibitemOpen
  \bibfield  {author} {\bibinfo {author} {\bibfnamefont {B.}~\bibnamefont
  {Smit}}\ and\ \bibinfo {author} {\bibfnamefont {D.}~\bibnamefont {Frenkel}},\
  }\href@noop {} {\bibfield  {journal} {\bibinfo  {journal} {Molecular
  physics}\ }\textbf {\bibinfo {volume} {68}},\ \bibinfo {pages} {951}
  (\bibinfo {year} {1989})}\BibitemShut {NoStop}%
\bibitem [{\citenamefont {Varnik}\ \emph {et~al.}(2000)\citenamefont {Varnik},
  \citenamefont {Baschnagel},\ and\ \citenamefont
  {Binder}}]{varnik2000molecular}%
  \BibitemOpen
  \bibfield  {author} {\bibinfo {author} {\bibfnamefont {F.}~\bibnamefont
  {Varnik}}, \bibinfo {author} {\bibfnamefont {J.}~\bibnamefont {Baschnagel}},
  \ and\ \bibinfo {author} {\bibfnamefont {K.}~\bibnamefont {Binder}},\
  }\href@noop {} {\bibfield  {journal} {\bibinfo  {journal} {The Journal of
  Chemical Physics}\ }\textbf {\bibinfo {volume} {113}},\ \bibinfo {pages}
  {4444} (\bibinfo {year} {2000})}\BibitemShut {NoStop}%
\bibitem [{\citenamefont {Shi}\ \emph {et~al.}(2023)\citenamefont {Shi},
  \citenamefont {Smith}, \citenamefont {Santiso},\ and\ \citenamefont
  {Gubbins}}]{Shi2023}%
  \BibitemOpen
  \bibfield  {author} {\bibinfo {author} {\bibfnamefont {K.}~\bibnamefont
  {Shi}}, \bibinfo {author} {\bibfnamefont {E.~R.}\ \bibnamefont {Smith}},
  \bibinfo {author} {\bibfnamefont {E.~E.}\ \bibnamefont {Santiso}}, \ and\
  \bibinfo {author} {\bibfnamefont {K.~E.}\ \bibnamefont {Gubbins}},\ }\href
  {\doibase 10.1063/5.0132487} {\bibfield  {journal} {\bibinfo  {journal} {The
  Journal of Chemical Physics}\ }\textbf {\bibinfo {volume} {158}},\ \bibinfo
  {pages} {040901} (\bibinfo {year} {2023})}\BibitemShut {NoStop}%
\bibitem [{\citenamefont {Marchio}\ \emph {et~al.}(2018)\citenamefont
  {Marchio}, \citenamefont {Meloni}, \citenamefont {Giacomello}, \citenamefont
  {Valeriani},\ and\ \citenamefont {Casciola}}]{marchio2018pressure}%
  \BibitemOpen
  \bibfield  {author} {\bibinfo {author} {\bibfnamefont {S.}~\bibnamefont
  {Marchio}}, \bibinfo {author} {\bibfnamefont {S.}~\bibnamefont {Meloni}},
  \bibinfo {author} {\bibfnamefont {A.}~\bibnamefont {Giacomello}}, \bibinfo
  {author} {\bibfnamefont {C.}~\bibnamefont {Valeriani}}, \ and\ \bibinfo
  {author} {\bibfnamefont {C.}~\bibnamefont {Casciola}},\ }\href@noop {}
  {\bibfield  {journal} {\bibinfo  {journal} {The Journal of chemical physics}\
  }\textbf {\bibinfo {volume} {148}},\ \bibinfo {pages} {064706} (\bibinfo
  {year} {2018})}\BibitemShut {NoStop}%
\bibitem [{\citenamefont {Cheng}(2022)}]{Cheng2022computing}%
  \BibitemOpen
  \bibfield  {author} {\bibinfo {author} {\bibfnamefont {B.}~\bibnamefont
  {Cheng}},\ }\href {\doibase 10.1063/5.0107059} {\bibfield  {journal}
  {\bibinfo  {journal} {The Journal of Chemical Physics}\ }\textbf {\bibinfo
  {volume} {157}},\ \bibinfo {pages} {121101} (\bibinfo {year}
  {2022})}\BibitemShut {NoStop}%
\bibitem [{\citenamefont {Li}\ \emph {et~al.}(1999)\citenamefont {Li},
  \citenamefont {Eddaoudi}, \citenamefont {O'Keeffe},\ and\ \citenamefont
  {Yaghi}}]{yaghi1999MOF5}%
  \BibitemOpen
  \bibfield  {author} {\bibinfo {author} {\bibfnamefont {H.}~\bibnamefont
  {Li}}, \bibinfo {author} {\bibfnamefont {M.}~\bibnamefont {Eddaoudi}},
  \bibinfo {author} {\bibfnamefont {M.}~\bibnamefont {O'Keeffe}}, \ and\
  \bibinfo {author} {\bibfnamefont {O.~M.}\ \bibnamefont {Yaghi}},\ }\href@noop
  {} {\bibfield  {journal} {\bibinfo  {journal} {Nature}\ }\textbf {\bibinfo
  {volume} {402}},\ \bibinfo {pages} {276} (\bibinfo {year}
  {1999})}\BibitemShut {NoStop}%
\bibitem [{\citenamefont {Frenkel}\ and\ \citenamefont
  {Ladd}(1984)}]{frenkel1984new}%
  \BibitemOpen
  \bibfield  {author} {\bibinfo {author} {\bibfnamefont {D.}~\bibnamefont
  {Frenkel}}\ and\ \bibinfo {author} {\bibfnamefont {A.~J.}\ \bibnamefont
  {Ladd}},\ }\href@noop {} {\bibfield  {journal} {\bibinfo  {journal} {The
  Journal of chemical physics}\ }\textbf {\bibinfo {volume} {81}},\ \bibinfo
  {pages} {3188} (\bibinfo {year} {1984})}\BibitemShut {NoStop}%
\bibitem [{\citenamefont {Cheng}\ and\ \citenamefont
  {Ceriotti}(2018)}]{cheng2018computing}%
  \BibitemOpen
  \bibfield  {author} {\bibinfo {author} {\bibfnamefont {B.}~\bibnamefont
  {Cheng}}\ and\ \bibinfo {author} {\bibfnamefont {M.}~\bibnamefont
  {Ceriotti}},\ }\href@noop {} {\bibfield  {journal} {\bibinfo  {journal}
  {Physical Review B}\ }\textbf {\bibinfo {volume} {97}},\ \bibinfo {pages}
  {054102} (\bibinfo {year} {2018})}\BibitemShut {NoStop}%
\bibitem [{\citenamefont {Cheng}\ and\ \citenamefont
  {Ceriotti}(2014)}]{cheng2014direct}%
  \BibitemOpen
  \bibfield  {author} {\bibinfo {author} {\bibfnamefont {B.}~\bibnamefont
  {Cheng}}\ and\ \bibinfo {author} {\bibfnamefont {M.}~\bibnamefont
  {Ceriotti}},\ }\href@noop {} {\bibfield  {journal} {\bibinfo  {journal} {The
  Journal of chemical physics}\ }\textbf {\bibinfo {volume} {141}},\ \bibinfo
  {pages} {244112} (\bibinfo {year} {2014})}\BibitemShut {NoStop}%
\bibitem [{\citenamefont {Li}\ \emph {et~al.}(2009)\citenamefont {Li},
  \citenamefont {Kuppler},\ and\ \citenamefont {Zhou}}]{RevMOFAds_Zhou2009}%
  \BibitemOpen
  \bibfield  {author} {\bibinfo {author} {\bibfnamefont {J.-R.}\ \bibnamefont
  {Li}}, \bibinfo {author} {\bibfnamefont {R.~J.}\ \bibnamefont {Kuppler}}, \
  and\ \bibinfo {author} {\bibfnamefont {H.-C.}\ \bibnamefont {Zhou}},\
  }\href@noop {} {\bibfield  {journal} {\bibinfo  {journal} {Chem. Soc. Rev.}\
  }\textbf {\bibinfo {volume} {38}},\ \bibinfo {pages} {1477} (\bibinfo {year}
  {2009})}\BibitemShut {NoStop}%
\bibitem [{\citenamefont {Getman}\ \emph {et~al.}(2012)\citenamefont {Getman},
  \citenamefont {Bae}, \citenamefont {Wilmer},\ and\ \citenamefont
  {Snurr}}]{RevMOFads_Snurr2012}%
  \BibitemOpen
  \bibfield  {author} {\bibinfo {author} {\bibfnamefont {R.~B.}\ \bibnamefont
  {Getman}}, \bibinfo {author} {\bibfnamefont {Y.-S.}\ \bibnamefont {Bae}},
  \bibinfo {author} {\bibfnamefont {C.~E.}\ \bibnamefont {Wilmer}}, \ and\
  \bibinfo {author} {\bibfnamefont {R.~Q.}\ \bibnamefont {Snurr}},\ }\href@noop
  {} {\bibfield  {journal} {\bibinfo  {journal} {Chem. Rev.}\ }\textbf
  {\bibinfo {volume} {112}},\ \bibinfo {pages} {703} (\bibinfo {year}
  {2012})}\BibitemShut {NoStop}%
\bibitem [{\citenamefont {Furukawa}\ \emph {et~al.}(2013)\citenamefont
  {Furukawa}, \citenamefont {Cordova}, \citenamefont {O’Keeffe},\ and\
  \citenamefont {Yaghi}}]{RevMOFAds_Yaghi2013}%
  \BibitemOpen
  \bibfield  {author} {\bibinfo {author} {\bibfnamefont {H.}~\bibnamefont
  {Furukawa}}, \bibinfo {author} {\bibfnamefont {K.~E.}\ \bibnamefont
  {Cordova}}, \bibinfo {author} {\bibfnamefont {M.}~\bibnamefont {O’Keeffe}},
  \ and\ \bibinfo {author} {\bibfnamefont {O.~M.}\ \bibnamefont {Yaghi}},\
  }\href@noop {} {\bibfield  {journal} {\bibinfo  {journal} {Science}\ }\textbf
  {\bibinfo {volume} {341}},\ \bibinfo {pages} {1230444} (\bibinfo {year}
  {2013})}\BibitemShut {NoStop}%
\bibitem [{\citenamefont {Zhou}\ \emph {et~al.}(2012)\citenamefont {Zhou},
  \citenamefont {Long},\ and\ \citenamefont {Yaghi}}]{zhou2012introduction}%
  \BibitemOpen
  \bibfield  {author} {\bibinfo {author} {\bibfnamefont {H.-C.}\ \bibnamefont
  {Zhou}}, \bibinfo {author} {\bibfnamefont {J.~R.}\ \bibnamefont {Long}}, \
  and\ \bibinfo {author} {\bibfnamefont {O.~M.}\ \bibnamefont {Yaghi}},\
  }\href@noop {} {\enquote {\bibinfo {title} {Introduction to metal--organic
  frameworks},}\ } (\bibinfo {year} {2012})\BibitemShut {NoStop}%
\bibitem [{\citenamefont {Sumida}\ \emph {et~al.}(2012)\citenamefont {Sumida},
  \citenamefont {Rogow}, \citenamefont {Mason}, \citenamefont {McDonald},
  \citenamefont {Bloch}, \citenamefont {Herm}, \citenamefont {Bae},\ and\
  \citenamefont {Long}}]{sumida2012carbon}%
  \BibitemOpen
  \bibfield  {author} {\bibinfo {author} {\bibfnamefont {K.}~\bibnamefont
  {Sumida}}, \bibinfo {author} {\bibfnamefont {D.~L.}\ \bibnamefont {Rogow}},
  \bibinfo {author} {\bibfnamefont {J.~A.}\ \bibnamefont {Mason}}, \bibinfo
  {author} {\bibfnamefont {T.~M.}\ \bibnamefont {McDonald}}, \bibinfo {author}
  {\bibfnamefont {E.~D.}\ \bibnamefont {Bloch}}, \bibinfo {author}
  {\bibfnamefont {Z.~R.}\ \bibnamefont {Herm}}, \bibinfo {author}
  {\bibfnamefont {T.-H.}\ \bibnamefont {Bae}}, \ and\ \bibinfo {author}
  {\bibfnamefont {J.~R.}\ \bibnamefont {Long}},\ }\href@noop {} {\bibfield
  {journal} {\bibinfo  {journal} {Chemical reviews}\ }\textbf {\bibinfo
  {volume} {112}},\ \bibinfo {pages} {724} (\bibinfo {year}
  {2012})}\BibitemShut {NoStop}%
\bibitem [{\citenamefont {Walton}\ \emph {et~al.}(2007)\citenamefont {Walton},
  \citenamefont {Millward}, \citenamefont {Dubbeldam}, \citenamefont {Frost},
  \citenamefont {Low}, \citenamefont {Yaghi},\ and\ \citenamefont
  {Snurr}}]{Walton2007}%
  \BibitemOpen
  \bibfield  {author} {\bibinfo {author} {\bibfnamefont {K.~S.}\ \bibnamefont
  {Walton}}, \bibinfo {author} {\bibfnamefont {A.~R.}\ \bibnamefont
  {Millward}}, \bibinfo {author} {\bibfnamefont {D.}~\bibnamefont {Dubbeldam}},
  \bibinfo {author} {\bibfnamefont {H.}~\bibnamefont {Frost}}, \bibinfo
  {author} {\bibfnamefont {J.~J.}\ \bibnamefont {Low}}, \bibinfo {author}
  {\bibfnamefont {O.~M.}\ \bibnamefont {Yaghi}}, \ and\ \bibinfo {author}
  {\bibfnamefont {R.~Q.}\ \bibnamefont {Snurr}},\ }\href {\doibase
  10.1021/ja076595g} {\bibfield  {journal} {\bibinfo  {journal} {Journal of the
  American Chemical Society}\ }\textbf {\bibinfo {volume} {130}},\ \bibinfo
  {pages} {406} (\bibinfo {year} {2007})}\BibitemShut {NoStop}%
\bibitem [{\citenamefont {Bureekaew}\ \emph {et~al.}(2013)\citenamefont
  {Bureekaew}, \citenamefont {Amirjalayer}, \citenamefont {Tafipolsky},
  \citenamefont {Spickermann}, \citenamefont {Roy},\ and\ \citenamefont
  {Schmid}}]{PSSB2013MOFFF}%
  \BibitemOpen
  \bibfield  {author} {\bibinfo {author} {\bibfnamefont {S.}~\bibnamefont
  {Bureekaew}}, \bibinfo {author} {\bibfnamefont {S.}~\bibnamefont
  {Amirjalayer}}, \bibinfo {author} {\bibfnamefont {M.}~\bibnamefont
  {Tafipolsky}}, \bibinfo {author} {\bibfnamefont {C.}~\bibnamefont
  {Spickermann}}, \bibinfo {author} {\bibfnamefont {T.~K.}\ \bibnamefont
  {Roy}}, \ and\ \bibinfo {author} {\bibfnamefont {R.}~\bibnamefont {Schmid}},\
  }\href@noop {} {\bibfield  {journal} {\bibinfo  {journal} {physica status
  solidi (b)}\ }\textbf {\bibinfo {volume} {250}},\ \bibinfo {pages} {1128}
  (\bibinfo {year} {2013})}\BibitemShut {NoStop}%
\bibitem [{\citenamefont {Keupp}\ \emph {et~al.}(2021)\citenamefont {Keupp},
  \citenamefont {Dürholt},\ and\ \citenamefont {Schmid}}]{Keupp2021FD}%
  \BibitemOpen
  \bibfield  {author} {\bibinfo {author} {\bibfnamefont {J.}~\bibnamefont
  {Keupp}}, \bibinfo {author} {\bibfnamefont {J.~P.}\ \bibnamefont {Dürholt}},
  \ and\ \bibinfo {author} {\bibfnamefont {R.}~\bibnamefont {Schmid}},\
  }\href@noop {} {\bibfield  {journal} {\bibinfo  {journal} {Faraday Discuss.}\
  }\textbf {\bibinfo {volume} {225}},\ \bibinfo {pages} {324} (\bibinfo {year}
  {2021})}\BibitemShut {NoStop}%
\bibitem [{\citenamefont {Bussi}\ \emph {et~al.}(2007)\citenamefont {Bussi},
  \citenamefont {Donadio},\ and\ \citenamefont {Parrinello}}]{Bussi2007}%
  \BibitemOpen
  \bibfield  {author} {\bibinfo {author} {\bibfnamefont {G.}~\bibnamefont
  {Bussi}}, \bibinfo {author} {\bibfnamefont {D.}~\bibnamefont {Donadio}}, \
  and\ \bibinfo {author} {\bibfnamefont {M.}~\bibnamefont {Parrinello}},\
  }\href {\doibase 10.1063/1.2408420} {\bibfield  {journal} {\bibinfo
  {journal} {J.\ Chem.\ Phys.}\ }\textbf {\bibinfo {volume} {126}},\ \bibinfo
  {pages} {14101} (\bibinfo {year} {2007})}\BibitemShut {NoStop}%
\bibitem [{\citenamefont {Plimpton}(1995)}]{Plimpton1995}%
  \BibitemOpen
  \bibfield  {author} {\bibinfo {author} {\bibfnamefont {S.}~\bibnamefont
  {Plimpton}},\ }\href {\doibase 10.1006/jcph.1995.1039} {\bibfield  {journal}
  {\bibinfo  {journal} {J.\ Comput.\ Phys.}\ }\textbf {\bibinfo {volume}
  {117}},\ \bibinfo {pages} {1} (\bibinfo {year} {1995})}\BibitemShut {NoStop}%
\bibitem [{\citenamefont {Potoff}\ and\ \citenamefont
  {Siepmann}(2001)}]{TraPPE_CO2}%
  \BibitemOpen
  \bibfield  {author} {\bibinfo {author} {\bibfnamefont {J.~J.}\ \bibnamefont
  {Potoff}}\ and\ \bibinfo {author} {\bibfnamefont {J.~I.}\ \bibnamefont
  {Siepmann}},\ }\href@noop {} {\bibfield  {journal} {\bibinfo  {journal}
  {AIChE Journal}\ }\textbf {\bibinfo {volume} {47}},\ \bibinfo {pages} {1676}
  (\bibinfo {year} {2001})}\BibitemShut {NoStop}%
\bibitem [{\citenamefont {Alexiadis}\ and\ \citenamefont
  {Kassinos}(2008)}]{alexiadis2008molecular}%
  \BibitemOpen
  \bibfield  {author} {\bibinfo {author} {\bibfnamefont {A.}~\bibnamefont
  {Alexiadis}}\ and\ \bibinfo {author} {\bibfnamefont {S.}~\bibnamefont
  {Kassinos}},\ }\href@noop {} {\bibfield  {journal} {\bibinfo  {journal}
  {Chemical reviews}\ }\textbf {\bibinfo {volume} {108}},\ \bibinfo {pages}
  {5014} (\bibinfo {year} {2008})}\BibitemShut {NoStop}%
\bibitem [{\citenamefont {Molinero}\ and\ \citenamefont
  {Moore}(2009)}]{molinero2009water}%
  \BibitemOpen
  \bibfield  {author} {\bibinfo {author} {\bibfnamefont {V.}~\bibnamefont
  {Molinero}}\ and\ \bibinfo {author} {\bibfnamefont {E.~B.}\ \bibnamefont
  {Moore}},\ }\href@noop {} {\bibfield  {journal} {\bibinfo  {journal} {The
  Journal of Physical Chemistry B}\ }\textbf {\bibinfo {volume} {113}},\
  \bibinfo {pages} {4008} (\bibinfo {year} {2009})}\BibitemShut {NoStop}%
\bibitem [{\citenamefont {Davies}\ \emph {et~al.}(2021)\citenamefont {Davies},
  \citenamefont {Fitzner},\ and\ \citenamefont
  {Michaelides}}]{davies2021routes}%
  \BibitemOpen
  \bibfield  {author} {\bibinfo {author} {\bibfnamefont {M.~B.}\ \bibnamefont
  {Davies}}, \bibinfo {author} {\bibfnamefont {M.}~\bibnamefont {Fitzner}}, \
  and\ \bibinfo {author} {\bibfnamefont {A.}~\bibnamefont {Michaelides}},\
  }\href@noop {} {\bibfield  {journal} {\bibinfo  {journal} {Proceedings of the
  National Academy of Sciences}\ }\textbf {\bibinfo {volume} {118}},\ \bibinfo
  {pages} {e2025245118} (\bibinfo {year} {2021})}\BibitemShut {NoStop}%
\bibitem [{\citenamefont {Werder}\ \emph {et~al.}(2003)\citenamefont {Werder},
  \citenamefont {Walther}, \citenamefont {Jaffe}, \citenamefont {Halicioglu},\
  and\ \citenamefont {Koumoutsakos}}]{werder2003water}%
  \BibitemOpen
  \bibfield  {author} {\bibinfo {author} {\bibfnamefont {T.}~\bibnamefont
  {Werder}}, \bibinfo {author} {\bibfnamefont {J.~H.}\ \bibnamefont {Walther}},
  \bibinfo {author} {\bibfnamefont {R.}~\bibnamefont {Jaffe}}, \bibinfo
  {author} {\bibfnamefont {T.}~\bibnamefont {Halicioglu}}, \ and\ \bibinfo
  {author} {\bibfnamefont {P.}~\bibnamefont {Koumoutsakos}},\ }\href@noop {}
  {\bibfield  {journal} {\bibinfo  {journal} {The Journal of Physical Chemistry
  B}\ }\textbf {\bibinfo {volume} {107}},\ \bibinfo {pages} {1345} (\bibinfo
  {year} {2003})}\BibitemShut {NoStop}%
\bibitem [{\citenamefont {Striolo}\ \emph {et~al.}(2005)\citenamefont
  {Striolo}, \citenamefont {Chialvo}, \citenamefont {Gubbins},\ and\
  \citenamefont {Cummings}}]{Striolo2005}%
  \BibitemOpen
  \bibfield  {author} {\bibinfo {author} {\bibfnamefont {A.}~\bibnamefont
  {Striolo}}, \bibinfo {author} {\bibfnamefont {A.~A.}\ \bibnamefont
  {Chialvo}}, \bibinfo {author} {\bibfnamefont {K.~E.}\ \bibnamefont
  {Gubbins}}, \ and\ \bibinfo {author} {\bibfnamefont {P.~T.}\ \bibnamefont
  {Cummings}},\ }\href {\doibase 10.1063/1.1924697} {\bibfield  {journal}
  {\bibinfo  {journal} {The Journal of Chemical Physics}\ }\textbf {\bibinfo
  {volume} {122}},\ \bibinfo {pages} {234712} (\bibinfo {year}
  {2005})}\BibitemShut {NoStop}%
\bibitem [{\citenamefont {Wang}\ \emph {et~al.}(2020)\citenamefont {Wang},
  \citenamefont {Wang}, \citenamefont {Li}, \citenamefont {Zheng},
  \citenamefont {Guan}, \citenamefont {Huang}, \citenamefont {Xu},\ and\
  \citenamefont {Yu}}]{wang2020porous}%
  \BibitemOpen
  \bibfield  {author} {\bibinfo {author} {\bibfnamefont {H.}~\bibnamefont
  {Wang}}, \bibinfo {author} {\bibfnamefont {X.}~\bibnamefont {Wang}}, \bibinfo
  {author} {\bibfnamefont {M.}~\bibnamefont {Li}}, \bibinfo {author}
  {\bibfnamefont {L.}~\bibnamefont {Zheng}}, \bibinfo {author} {\bibfnamefont
  {D.}~\bibnamefont {Guan}}, \bibinfo {author} {\bibfnamefont {X.}~\bibnamefont
  {Huang}}, \bibinfo {author} {\bibfnamefont {J.}~\bibnamefont {Xu}}, \ and\
  \bibinfo {author} {\bibfnamefont {J.}~\bibnamefont {Yu}},\ }\href@noop {}
  {\bibfield  {journal} {\bibinfo  {journal} {Advanced materials}\ }\textbf
  {\bibinfo {volume} {32}},\ \bibinfo {pages} {2002559} (\bibinfo {year}
  {2020})}\BibitemShut {NoStop}%
\bibitem [{\citenamefont {Slater}\ and\ \citenamefont
  {Cooper}(2015)}]{slater2015function}%
  \BibitemOpen
  \bibfield  {author} {\bibinfo {author} {\bibfnamefont {A.~G.}\ \bibnamefont
  {Slater}}\ and\ \bibinfo {author} {\bibfnamefont {A.~I.}\ \bibnamefont
  {Cooper}},\ }\href@noop {} {\bibfield  {journal} {\bibinfo  {journal}
  {Science}\ }\textbf {\bibinfo {volume} {348}},\ \bibinfo {pages} {aaa8075}
  (\bibinfo {year} {2015})}\BibitemShut {NoStop}%
\bibitem [{\citenamefont {Adebajo}\ \emph {et~al.}(2003)\citenamefont
  {Adebajo}, \citenamefont {Frost}, \citenamefont {Kloprogge}, \citenamefont
  {Carmody},\ and\ \citenamefont {Kokot}}]{adebajo2003porous}%
  \BibitemOpen
  \bibfield  {author} {\bibinfo {author} {\bibfnamefont {M.~O.}\ \bibnamefont
  {Adebajo}}, \bibinfo {author} {\bibfnamefont {R.~L.}\ \bibnamefont {Frost}},
  \bibinfo {author} {\bibfnamefont {J.~T.}\ \bibnamefont {Kloprogge}}, \bibinfo
  {author} {\bibfnamefont {O.}~\bibnamefont {Carmody}}, \ and\ \bibinfo
  {author} {\bibfnamefont {S.}~\bibnamefont {Kokot}},\ }\href@noop {}
  {\bibfield  {journal} {\bibinfo  {journal} {Journal of Porous materials}\
  }\textbf {\bibinfo {volume} {10}},\ \bibinfo {pages} {159} (\bibinfo {year}
  {2003})}\BibitemShut {NoStop}%
\bibitem [{\citenamefont {Liu}\ \emph {et~al.}(2022)\citenamefont {Liu},
  \citenamefont {Verma}, \citenamefont {Chen}, \citenamefont {Hu},
  \citenamefont {Huang}, \citenamefont {Yang}, \citenamefont {Ma},\ and\
  \citenamefont {Wang}}]{liu2022metal}%
  \BibitemOpen
  \bibfield  {author} {\bibinfo {author} {\bibfnamefont {X.}~\bibnamefont
  {Liu}}, \bibinfo {author} {\bibfnamefont {G.}~\bibnamefont {Verma}}, \bibinfo
  {author} {\bibfnamefont {Z.}~\bibnamefont {Chen}}, \bibinfo {author}
  {\bibfnamefont {B.}~\bibnamefont {Hu}}, \bibinfo {author} {\bibfnamefont
  {Q.}~\bibnamefont {Huang}}, \bibinfo {author} {\bibfnamefont
  {H.}~\bibnamefont {Yang}}, \bibinfo {author} {\bibfnamefont {S.}~\bibnamefont
  {Ma}}, \ and\ \bibinfo {author} {\bibfnamefont {X.}~\bibnamefont {Wang}},\
  }\href@noop {} {\bibfield  {journal} {\bibinfo  {journal} {The Innovation}\
  ,\ \bibinfo {pages} {100281}} (\bibinfo {year} {2022})}\BibitemShut {NoStop}%
\end{thebibliography}
\end{document}